\documentclass[twocolumn,amsmath,amssymb,superscriptaddress,footinbib]{revtex4-2}

\usepackage{graphicx}% Include figure files
\usepackage{dcolumn}% Align table columns on decimal point
\usepackage{bm}% bold math
\usepackage{braket}
\usepackage{hyperref}

\begin{document}

\title{Structures resistant to Manipulation by \textit{all} Wavefronts in two dimensions}
%\title{Structures resistant to Manipulation by \textit{all} Wavefronts in 2D}

\newcommand{\MITphys}{Department of Physics, Massachusetts Institute of Technology, Cambridge, MA 02139, USA.}
\newcommand{\TUW}{Photonics Institute, Vienna University of Technology (TU Wien), Vienna, A-1040, Austria}
\newcommand{\MITRLE}{Research Laboratory of Electronics, Massachusetts Institute of Technology, Cambridge, MA 02139, USA.}

\author{Asher Sabbagh}
\affiliation{\MITphys}
\author{Michael Horodynski}
\email{michael.horodynski@tuwien.ac.at}
\affiliation{\MITphys}
\affiliation{\TUW}
\author{Rida Khan}
\affiliation{\MITphys}
\author{Brian Shi}
\affiliation{\MITphys}
\author{Marin Solja\v{c}i\'{c}}
\affiliation{\MITphys}
\affiliation{\MITRLE}

\begin{abstract}
    Using light to manipulate small particles is a powerful tool with numerous practical applications across biophysics and nanotechnology. This experimental technique has achieved significant performance gains by employing shaped wavefronts, most commonly generated with spatial light modulators. Wavefront shaping has also enabled the manipulation of seemingly arbitrary objects beyond the reach of conventional beams. Contrary to this established assumption, we show here the existence of a wide variety of objects resistant to manipulation, even with the optimal wavefront shaping protocol. The counterintuitive shapes of these objects are found using inverse design in two dimensions, providing a foundation for their natural extension to three dimensions. Specifically, we show that the maximal pulling force is reduced by up to four orders of magnitude, and the maximal trapping stiffness is reduced by up to nearly two orders of magnitude. Our findings could prove useful for the development of micromachines that require a predictable mechanical response to arbitrary waves.
\end{abstract}

\maketitle

\subsection*{Introduction}

Since the pioneering work of Ashkin, who first demonstrated that tightly focused light can stably trap microscopic particles \cite{ashkin_observation_1986}, optical micromanipulation has developed into a versatile tool across physics, biology, and soft matter \cite{padgett_tweezers_2011,dholakia_shaping_2011,volpe_roadmap_2023}. For example, optical trapping has been used to elucidate the biomechanics of biological processes \cite{block_bead_1990,heller_optical_2014}, control micromachines \cite{grier_revolution_2003,butaite_indirect_2019}, experimentally validate the connection between entropy and information \cite{berut_experimental_2012,saha_maximizing_2021}, and aid in cooling nanoparticles to their motional ground state \cite{delic_cooling_2020,tebbenjohanns_quantum_2021}. The same optical fields can also transfer angular momentum \cite{friese_optical_1998} and even generate pulling forces \cite{novitsky_single_2011,brzobohaty_experimental_2013}. Related capabilities also exist in acoustics, including acoustic tweezers for manipulating single microparticles, cells, and entire organisms \cite{ding_-chip_2012} and acoustic tractor beams \cite{demore_acoustic_2014}.

The ability to shape the wavefront of light \cite{mosk_controlling_2012,rotter_light_2017,forbes_structured_2021,bertolotti_imaging_2022} beyond the conventionally used Gaussian beams has advanced micromanipulation capabilities \cite{marzo_holographic_2015,otte_optical_2020}. For example, wavefront shaping enables trapping behind scattering layers \cite{cizmar_situ_2010}, stiffer traps \cite{taylor_enhanced_2015,butaite_photon-efficient_2024}, controlled transfer of angular momentum \cite{stilgoe_controlled_2022}, trapping at multiple spots \cite{garces-chavez_simultaneous_2002}, and trapping and manipulating rod-shaped bacteria \cite{lenton_orientation_2020}. Alternatively, the near-field around the particle can be tailored by inverse-designing nanocavities and plasmonic apertures \cite{martinez_de_aguirre_jokisch_omnidirectional_2024,nelson_inverse_2024}. Acoustic transducer arrays provide an analogous form of control and can, for example, drive objects even through disordered and dynamic media \cite{orazbayev_wave-momentum_2024}. A complementary route for improving micromanipulation performance is inverse-designing particles for fixed wavefront illumination (commonly a plane wave or Gaussian beam) \cite{phillips_shape-induced_2014,stein_shaping_2022,lee_inverse_2025}. Bounds on optimal forces, torques, and illumination fields \cite{liu_optimal_2019,kuang_maximal_2020} solidified the understanding that shaped waves have a large influence for manipulating objects. These successes raise a fundamental question: are there targets for which mechanical control remains fundamentally limited even with optimal wavefront shaping?

In this paper, we identify targets for which even the optimal incident wavefront cannot efficiently realize certain micromanipulation tasks. We inverse-design structures whose mechanical response is predictable for an arbitrary wavefront (within fixed illumination constraints). As one example, we numerically show that the maximum achievable pulling force (``optimal tractor beam'') on these inverse-designed targets can be suppressed by up to four orders of magnitude relative to a generic structure with a uniform refractive index distribution. Remarkably, the ability to design such structures persists across broad parameter ranges, indicating that wavefront shaping does not guarantee arbitrary mechanical control. Our results not only show that it is possible to find structures that are predictably manipulated by an arbitrary wave, enabling reliable micromanipulation with dynamically distorted wavefronts. They also demonstrate fundamental limits to improving micromanipulation using wavefront shaping.

\begin{figure}[t!]
    \includegraphics[width=\linewidth]{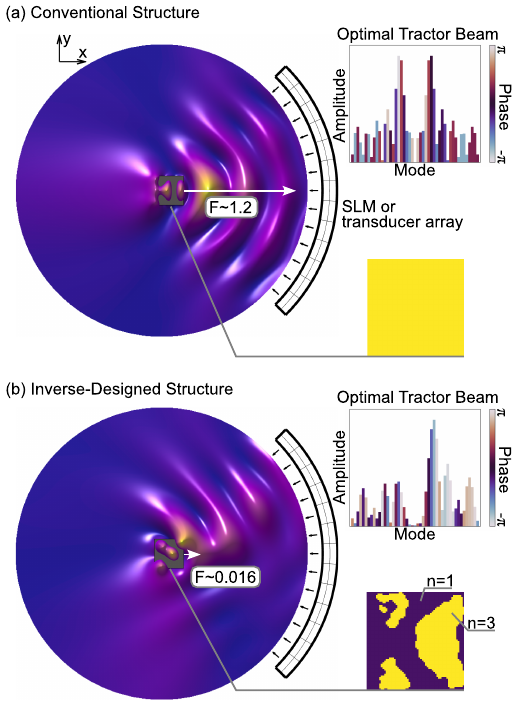}
    \caption{\textbf{Reduction of optimal tractor beam strength with inverse design.} Intensity distribution resulting from the optimal tractor beam for a conventional structure (a) with a uniform refractive index distribution ($n=3$) and for an inverse-designed structure (b) to minimize the optimal pulling force. Here, the maximal pulling force ($F$) on the inverse-designed structure is reduced by nearly two orders of magnitude compared to the conventional structure. We restrict the aperture of the incoming wave to one-quarter of the circular boundary. Insets display the refractive index distributions of both structures and the modal distributions of the optimal tractor beam wavefronts. The gray square indicates the extent of the design region.} 
    \label{fig:1}
\end{figure}

\begin{figure*}[t!]
    \includegraphics{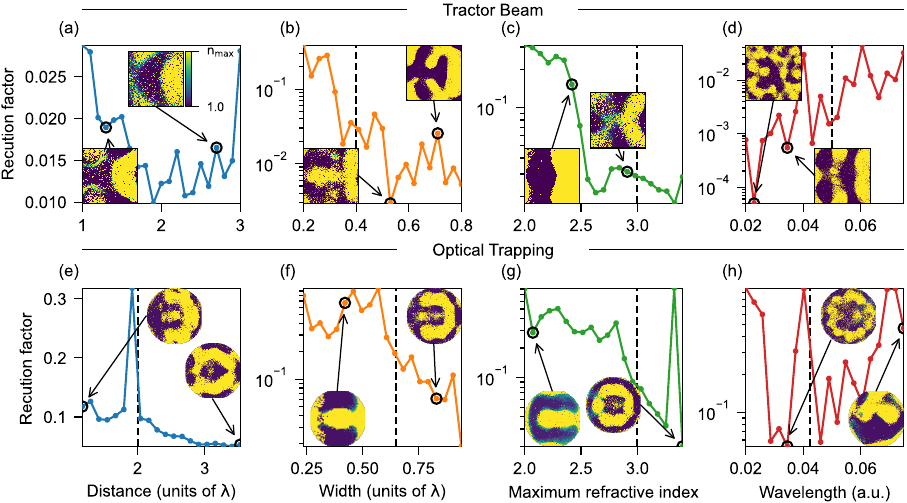}
    \caption{\textbf{Broad parameter range for tractor beam and trapping resistant particles.} Plots show by how much the optimal micromanipulation has been reduced for an inverse-designed target compared to a conventional structure with a uniform refractive index. The top row and bottom give the reduction factor for the optimal tractor beam (i.e., the largest eigenvalue of $Q_x$) and optimal trapping stiffness (i.e., the largest eigenvalue of $W_x$), respectively. The reduction factor is shown as a function of distance (a, e), design-region width (b, f), maximal refractive index of the structure (c, g), and wavelength (d, h). Insets show examples of inverse-designed structures that minimize this ratio. The black dashed line indicates the parameter value used in the other scans. The only exception is the wavelength scans, which are performed at a width of 0.04 a.u. to also investigate structures larger than a single wavelength.}
    \label{fig:2}
\end{figure*}

\subsection*{Inverse design of manipulation-resistant structures}

To inverse-design structures where the optimal wavefront exhibits a negligible influence on the target, we first need the ability to compute the optimal wavefront. For the first example of targets designed for greatly reduced optimal pulling forces, we employ the generalized Wigner-Smith operator $Q_x$, which was introduced for optimal focusing on targets in disordered media \cite{ambichl_focusing_2017} and has in the meantime been applied to a wide range of manipulation tasks \cite{horodynski_optimal_2020,horodynski_tractor_2023,butaite_photon-efficient_2024,hupfl_optimal_2023}. The optimal tractor beam wavefront is found by solving the eigenvalue problem of $Q_x$ (which is Hermitian in the absence of loss and gain in the system) for the largest eigenvalue, which reads
\begin{equation}
    Q_x \vec{u}_x = (K_x^{\text{in}}-S^{\dagger}K_x^{\text{out}}S) \vec{u}_x = q_x \vec{u}_x,
    \label{eq:momentum}
\end{equation}
where $S$ is the target's scattering matrix, $K_x^{\text{in}}$ ($K_x^{\text{out}}$) describes the incoming (outgoing) momentum of an arbitrary wavefront, $\vec{u}_x$ is a vector of modal amplitudes, and $q_x$ is the eigenvalue. From this equation, we can interpret $Q_x$ as describing the difference between the wave's incoming and outgoing momentum, which is transferred (by conservation of momentum) to the target. We restrict our analysis to two spatial dimensions, where the scattering of waves is governed by the Helmholtz equation $[\Delta+k^2n^2(\textbf{r})]\psi(\textbf{r})=0$, with $\Delta$ the Laplacian, $k=|\textbf{k}|$ the wavevector, $\psi(\textbf{r})$ the scalar wave, and $n(\textbf{r})$ is the spatially varying refractive index (which is continually adjusted during the inverse design process). We solve the two-dimensional Helmholtz equation numerically using our publicly available code \cite{horodynski_open_2023} based on an open-source finite-element method (NGSolve) \cite{schoberl_netgen_1997,schoberl_c11_2014}. We note that our approach can be straightforwardly extended to three dimensions and two polarizations (though at a higher computational cost).

Our ultimate goal is to find structures for which the largest eigenvalue of $Q_x$, $q_x^\text{max}$ is vanishingly small. Since this eigenvalue gives --- for all possible incoming wavefronts --- the largest pulling force (i.e., the strongest tractor beam), $q_x^\text{max} \sim 0$ implies that it is impossible to pull the target with \textit{any} wave. We achieve this goal by employing inverse design, in which we continuously adjust the refractive index distribution of the structure (parameterized by a square grid of pixels with individually tunable refractive indices between a minimum and maximum value) guided by a cost function (which is just $q_x^\text{max}$). To use state-of-the-art gradient-based inverse design algorithms, we require a computationally efficient way of calculating the gradient of the cost function with respect to the refractive index, $n_i$, of each design pixel, i.e., $\partial_{n_i}q_x^\text{max} = \left(\vec{u}_x^\text{max}\right)^\dagger \partial_{n_i}Q_x\vec{u}_x^\text{max}$. From Eq.~\eqref{eq:momentum}, we see that calculating $\partial_{n_i}Q_x$ amounts to the computation of $\partial_{n_i}S$, since $K_x$ is independent of the refractive index distribution. We efficiently calculate the sensitivity of the scattering matrix with respect to the refractive indices of the design pixels by using a method based on an alternative form of the generalized Wigner-Smith operator naturally suited to the task at hand \cite{horodynski_anti-reflection_2022}. We refer to the Supplementary Information for a detailed description of the gradient computation and inverse design algorithm.

In Fig.~\ref{fig:1}, we show the resulting structure of one such design run and compare its performance to that of a conventional structure. More specifically, we place a square design region ($0.4 \lambda \times 0.4\lambda$) with a grid of $60 \times 60$ design pixels in the middle of a circular scattering region. We can inject wavefronts into the scattering region with an arbitrary amplitude and phase distribution in a $\pi/2$ solid angle. These wavefronts always have unit norm for comparable results, i.e., $\vec{u}^\dagger \vec{u}=1$. Our simulations reveal an optimal pulling force on the conventional structure (a square with a uniform refractive index of $n=3$) of $F_x\sim 1.2$ (a.u.). In contrast, the optimal pulling force on the inverse-designed structure is smaller by nearly two orders of magnitude: $F_x\sim0.016$ (a.u.). This drop in pulling force is achieved despite comparable intensity to the baseline case in the design region. We note that the wave's ability to apply a pushing force is (in principle) unaffected by our design procedure. In this example, the maximal pushing force on the inverse-designed [conventional] target is $F\sim -89.4$ (a.u.) [$F\sim-84.2$ (a.u.)]. 

\subsection*{Tractor beam resistant structures}

Having presented one example of an inverse-designed structure where the optimal pulling force is reduced by two orders of magnitude, we now turn our attention to the question of whether such structures can be found generically. To this end, we vary the parameters of the studied system and investigate the extent to which the optimal pulling force is reduced relative to that of a corresponding conventional structure. Concretely, we sweep the distance of the structure to the wave's source, the side length of the square design region, the maximum refractive index, and the wavelength. We note that each structure is designed for one particular wavelength, i.e., for monochromatic illumination, and that changing the distance of the structure to the source also indicates the influence of varying the numerical aperture. 

Our results are summarized in Fig.~\ref{fig:2}a-d and clearly show that structures (strongly) resistant to the optimal pulling force occur over a wide range of parameters. This indicates that we have identified a large class of structures and that wavefront shaping does not, in general, guarantee arbitrary micromanipulation capabilities. 
Our data also clearly indicate that reducing the effective size of the design region (either by changing the size itself or indirectly by increasing the wavelength or decreasing the maximal refractive index) reduces the inverse-design algorithm's ability to identify structures that perform markedly better than conventional structures. Intuitively, this observation is explained by the transition to purely dipole-scattering for targets that are much smaller than the wavelength. In this case, the patterning is not resolved by the wave; thus, inverse design loses its influence on the optimal pulling force. These observations remain consistent if we restart the inverse design process with a different set of randomly selected refractive indices within the design region, indicating robustness to initialization. The wavelength scan is performed at a width of 0.04 a.u. to also include design regions up to $2\lambda \times 2\lambda$ in size. Even at these sizes, we can design structures resistant to manipulation by up to 4 orders of magnitude compared to a conventional structure. This is especially remarkable since, for large structures, wavefront shaping usually exhibits improved performance compared to smaller structures \cite{butaite_photon-efficient_2024,taylor_enhanced_2015}. %The most noteworthy deviation from this general trend was for a particular size ($0.53 \lambda$) of the design region, for which the performance was markedly worse than for comparable sizes. This spike appeared consistently, even after randomizing the initial guess of the design region's refractive index distribution three times per data point and running the inverse design process for more iterations. A possible explanation is a resonance effect, as the size in question is close to $\lambda/2$.

\begin{figure*}[t!]
    \centering
    \includegraphics{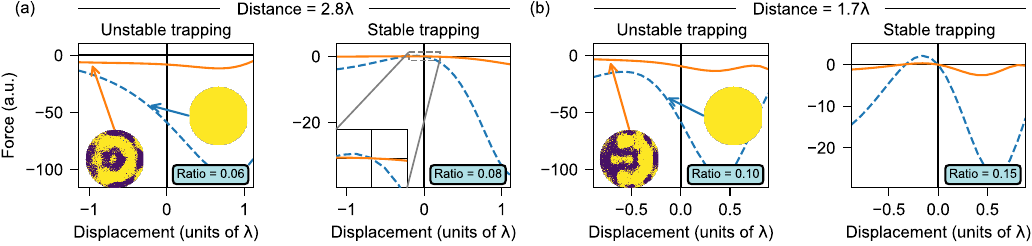}
    \caption{\textbf{Maximum trapping stiffness and stable trapping.} Force exerted as a function of displacement from the initial position for two different initial positions [(a) 2.8$\lambda$ and (b) 1.7$\lambda$, respectively]. In both cases, we compare the response of both the inverse-designed structure (for small optimal trapping stiffness) and the conventional structure with a uniform refractive index distribution. Furthermore, we apply two different wavefronts (left and right sides in a and b, respectively). One produces the optimal trapping stiffness as it corresponds to the largest eigenvalue of $W_x$, while the other maximizes the trapping stiffness under the constraint of stable trapping. Insets show the structures used and the ratio of the trapping stiffness between the inverse-designed and conventional structures by computing the ratio of the slopes at 0 displacement.}
    \label{fig:3}
\end{figure*}

\subsection*{Structures resistant to optical trapping}

We now show that the inverse design of structures resistant to manipulation also extends to the optical trapping of particles. Accordingly, we use an operator derived from the generalized Wigner-Smith operator in Eq.~\eqref{eq:momentum} that allows us to compute the wavefront with the optimal trapping stiffness. The trapping stiffness (in $x$-direction) $\kappa_x = -\partial_xF_x$ quantifies how tightly a particle is trapped by computing how quickly a restoring force builds when the particle is displaced from its equilibrium position. The corresponding operator is given by \cite{butaite_photon-efficient_2024}
\begin{equation}
    W_x = -\partial_x Q_x = i\left( S^\dagger K_x^\mathrm{out} S K_x^\mathrm{in} - K_x^\mathrm{in} S^\dagger K_x^\mathrm{out} S\right),
\end{equation}
where we used $\partial_xS=i (S K_x^\mathrm{in} - K_x^\mathrm{out} S)$ \cite{ambichl_focusing_2017} to arrive at $W_x$. 
Since the only part of $W_x$ that depends on the refractive index of the design pixels is the scattering matrix $S$, we employ the same inverse design strategy as for the tractor beam case. Here, we use a circular design region to connect our results to optical trapping experiments, which usually employ silica nanospheres as targets. This also means that the conventional structure used as the baseline is circular and has a uniform refractive-index distribution.

For structures designed to resist an optical trapping beam, we find a wide range of parameters in which the performance of the optimal trapping beam degrades by up to two orders of magnitude relative to a conventional structure (see Fig.~\ref{fig:2}e-h). The broad trend of larger structures performing better at resisting a trapping beam is similar to that observed for structures designed to withstand the optimal pulling force. Interestingly, with increasing distance to the wave's source, the performance of trapping-resistant targets improves, while the performance of tractor beam-resistant targets stays unchanged. In both micromanipulation examples considered here, the structures identified by our inverse-design process consist primarily of two materials, as we employed hyperbolic-tangent regularization to guide the optimizer toward solutions exhibiting this property. The structures also exhibit comparably smooth boundaries without nudging them towards this property in the inverse-design process.

However, in optical trapping, it is usually not sufficient to only target improved trapping stiffness with wavefront shaping \cite{volpe_roadmap_2023}. Rather, great care has to be taken to ensure that a trap is not only stiff, but also stable, i.e., no force is applied at zero displacement ($x=0$). In our case, this means that the trapping beam should not only maximize $\vec{u}^\dagger W_x \vec{u}$, but also ensure that $\vec{u}^\dagger Q_x \vec{u} = 0$ at the equilibrium position, where $\vec{u}$ is the optimal wavefront. We find wavestates that possess these two properties by numerical optimization (see SI for a detailed description of the procedure), as no simple eigenproblem is available to compute the corresponding wavefront \cite{butaite_photon-efficient_2024}.

We analyze the response of conventional and inverse-designed structures to wavefronts that maximize trapping stiffness while keeping the trap stable in Fig.~\ref{fig:3}. Our results reveal that changing the wavefront from one that just maximizes trapping stiffness to one that also results in stable trapping does not adversely affect the purpose of the inverse-designed structure. I.e., the optimal trapping stiffness is still reduced by approximately an order of magnitude, and the potential well created by the optical trap is much shallower for the inverse-designed structure.

\subsection*{Outlook and discussion}

To summarize, we have inverse-designed structures that are resistant to pulling forces and optical trapping beams, even when the \textit{optimal} wavefront for the task at hand is used. We illustrated the viability of our concept with numerical simulations, reducing the capability to manipulate inverse-designed structures by up to four orders of magnitude relative to conventional structures. The properties of targets we found potentially enable the application of our inverse-design pipeline to the fabrication of structures suited both for optical \cite{lee_inverse_2025}, and acoustic \cite{stein_shaping_2022} micromanipulation. Although the results presented here are for a two-dimensional system, the extension to three-dimensional vector waves should be possible.

Our approach offers an interesting perspective on optically driven micromachines \cite{zhang_fabrication_2022}, when only specific degrees of freedom need to be addressed. It potentially enables the operation of such micromachines even when they are embedded in a disordered, changing environment, i.e., when the illuminating wavefront is constantly evolving. Manipulation-resistant structures could also be used to reduce back-action noise in optical levers \cite{hao_back_2024}.

\subsection*{Acknowledgements}
This research was funded in part (M.H.) by the Austrian Science Fund (FWF) [10.55776/J4729]. This material is based upon work also supported in part by the U. S. Army Research Office through the Institute for Soldier Nanotechnologies at MIT, under Collaborative Agreement Number W911NF-23-2-0121. The authors acknowledge the MIT SuperCloud and Lincoln Laboratory Supercomputing Center for providing high-performance computing resources that have contributed to the research results reported within this paper. For open access purposes, the authors have applied a CC BY public copyright license to any author accepted manuscript version arising from this submission.

\bibliographystyle{apsrev4-2}
\bibliography{references.bib}

\end{document}

\title{Structures resistant to Manipulation by \textit{all} Wavefronts in two dimensions:\\ supplementary information}

\newcommand{\MITphys}{Department of Physics, Massachusetts Institute of Technology, Cambridge, MA 02139, USA.}
\newcommand{\TUW}{Photonics Institute, Vienna University of Technology (TU Wien), Vienna, A-1040, Austria}
\newcommand{\MITRLE}{Research Laboratory of Electronics, Massachusetts Institute of Technology, Cambridge, MA 02139, USA.}

\author{Asher Sabbagh}
\affiliation{\MITphys}
\author{Michael Horodynski}
\affiliation{\MITphys}
\affiliation{\TUW}
\author{Rida Khan}
\affiliation{\MITphys}
\author{Brian Shi}
\affiliation{\MITphys}
\author{Marin Solja\v{c}i\'{c}}
\affiliation{\MITphys}
\affiliation{\MITRLE}

\maketitle

\section{Scattering theory}
 Here, we recap the work of Ref.~\cite{horodynski_tractor_2023} on calculating the scattering matrix in two dimensions with open boundary conditions. Our starting point is the homogeneous Helmholtz equation:
\begin{equation}
    [\Delta + k^2n^2(\textbf{r})]\psi(\textbf{r})=0,
\end{equation}
where $\Delta$ is the Laplacian, $k=|\textbf{k}|$ is the wavevector, $n(\textbf{r})$ is the refractive index distribution, $\psi(\textbf{r})$ is the scalar wave, and $\textbf{r}$ is the position. Written in polar coordinates $(r, \varphi)$, the most general free space  solution is spanned by Hankel functions of the first and second kind:
\begin{equation}
    \psi(\textbf{r}) = \sum_{n \in \mathbb{Z}}{[\alpha_nH_n^{(2)}(kr)+\beta_nH_n^{(1)}(kr)](\gamma_ne^{in\varphi}+\delta_ne^{-in\varphi})},
    \label{eq:gensoln}
\end{equation}
where $\alpha_n, \beta_n, \gamma_n, \delta_n$ are the coefficients of the general expansion. This basis is complete and orthonormal, so we can use it to compute the matrix elements of a unitary scattering matrix, provided we restrict ourselves to flux-conserving systems (i.e., the absence of gain and loss). To arrive at a properly normalized expression for the scattering matrix, we consider the inhomogeneous Helmholtz equation with a source term $f(\textbf{r})$:
\begin{equation}
    [\Delta + k^2n^2(\textbf{r})]\psi(\textbf{r})=f(\textbf{r}).
\label{eq:inhomog}
\end{equation}
To excite an arbitrary incoming wave at the boundary of our design region $r=R$, we set $f(\textbf{r})=-\delta(r-R)\sum_n c_n e^{in\varphi}$, where the $c_n$'s are the modal amplitudes.  We then choose solutions inside and outside of the scattering region of the form
\begin{align}
    \psi_{\text{in}} & =\sum_n {a_n H^{(2)}_n(kr)e^{in\varphi}, r<R} \\
    \psi_{\text{out}}& =\sum_n {b_n H^{(1)}_n(kr)e^{in\varphi}, r>R},
\end{align}
where $a_n$ and $b_n$ are expansion coefficients.  We chose to study these forms of solutions because the norm of each component has a radial symmetry, which makes the source isotropic.  We also note that all components have constant flux, so we can easily construct a flux-conserving scattering matrix from them by using these solutions to fix the $c_n$'s from the source term \cite{horodynski_tractor_2023}.  To conserve the flux of our system, we enforce continuity of these solutions at $r=R$ to find that $a_nH_n^{(2)}(kR)=b_nH_n^{(1)}(kR)$. We can then plug these solutions back into Eq.~\eqref{eq:inhomog} to fix the source amplitudes, resulting in $c_n=-4ia_n / \pi R H_n^{(1)}(kR)$.  Note that Hankel functions of constant argument but increasing $n$ increase in their absolute value, so $|c_n|$ is decreasing with increasing $n$. This allows us to impose a heuristic cutoff to the number of modes, $n=N$, above which the modal amplitudes are effectively 0. Now that we know the full modal expansion of the source term, we can numerically compute the matrix elements of a unitary scattering matrix of an incoming cylindrical wavefront with unit modal amplitude:
\begin{equation}
    [S]_{mn}=\int_{0}^{2\pi}{\frac{e^{im\varphi}\psi_n(R)}{2\pi H_m^{(1)}(kR)}\mathrm{d}\varphi}-\frac{H_n^{(2)}(kR)}{H_n^{(1)}(kR)}\delta_{m,-n},
\label{eq:helmholtz}
\end{equation}
where the subtraction of the second term is to avoid double-counting the incoming wave's contribution. 

To compute the matrix elements of other relevant operators, we make use of the basis of incoming modes {${\chi}_n^\text{in}$}, by employing the fact that incoming waves are of the form $e^{in\varphi}H_n^{(2)}(kr)$ and normalizing.  The basis wavefunctions are given by
\begin{equation}
\chi_n^{\text{in}}=\frac{e^{in\varphi}H_n^{(2)}(kr)}{\sqrt{2\pi}|H_n^{(2)}(kR)|}.
    \label{eq:modes}
\end{equation}
We can then write out the matrix elements of the incoming momentum operator in the x-direction, $K_x^{\text{in}}$ evaluated at the boundary:
\begin{equation}
    [K^{\text{in}}_x]_{mn} =-i\int_0^{2\pi} \mathrm{d}\varphi (\chi^{\text{in}}_m)^* \frac{\partial}{\partial x} \chi^{\text{in}}_n,
\end{equation}
where we note that $K_x^{\text{in}}=K_x^{\text{out}}$.
By imposing our mode cutoff, the momentum operator is given by
\begin{equation}
    [K_x^{\text{in}}]_{mn}=\frac{ik}{4} \left[ \left( \frac{|H_{n}^{(1)}(kR)|}{|H_{m}^{(1)}(kR)|} + \frac{|H_{m}^{(1)}(kR)|}{|H_{n}^{(1)}(kR)|} \right) \delta_{m, n+1}- \left(\frac{|H_{n}^{(1)}(kR)|}{|H_{m}^{(1)}(kR)|} + \frac{|H_{m}^{(1)}(kR)|}{|H_{n}^{(1)}(kR)|} \right) \delta_{m+1, n} \right],
\end{equation}
where $m$ and $n$ run from $-N$ to $N$.  During our analysis, we restricted the incoming wavefront to an aperture of $\pi/2$. We achieve this by rewriting all relevant operators in the eigenbasis of the angular operator $\phi$, and then simply setting components outside of the aperture to zero. Again, we use the basis of incoming modes to compute the matrix elements of $\phi$:
\begin{equation}
    [\phi]_{mn} = \int_{0}^{2\pi}{(\chi_m^{\text{in}})^*\varphi\chi_n^{\text{in}}\mathrm{d}\varphi}.
\end{equation}

For Figure 3 of the main text, we need to compute the scattering matrix for a given structure at several positions. Rather than calculating the full scattering matrix from scratch at each of these points, which is computationally expensive, we compute the scattering matrix at a single position and then use the translation operator to evaluate it at any other position. This is the same translation operator used in standard quantum mechanics, generated by the momentum operator: $T(x)=e^{i  x K^{\text{in}}_x}$, $T^\dagger(x)=e^{-i  x K^{\text{out}}_x}$. To compute the scattering matrix at a distance $\Delta x$ from the starting position, we use:
\begin{equation}
    S(x+\Delta x) = T^\dagger(\Delta x)S(x)T(\Delta x).
\end{equation}

\section{Cost Function}

\subsection{Tractor Beam Micromanipulation}

To proceed with our inverse-design procedure, we must define a cost function and compute its gradient to optimize the refractive index distribution of our structure numerically. To generate structures that are resistant to a tractor beam, for example, we need a way to compute the force exerted onto the structure by the incoming wavefront and the gradient of that force under a change of the refractive index distribution. To do this, we introduce the formalism of the Generalized Wigner-Smith (GWS) operators, because they allow us to calculate all relevant quantities in terms of operators evaluated at the boundary, like $K_x^\text{in}$ and $S$. We define the GWS operator $Q_\alpha$ as \cite{horodynski_optimal_2020,ambichl_focusing_2017}
\begin{equation}
    Q_\alpha \equiv -iS^{\dagger}\frac{\partial}{\partial \alpha}S,
    \label{eq:qderiv}
\end{equation}
where $\alpha$ is some parameter of the system. We also define the scattering potential $U(\textbf{r}) \equiv -k^2n^2(\textbf{r})$. Then, the GWS operator satisfies the following equation, which relates far-field information at the boundary to the optimization region \cite{horodynski_optimal_2020}:
\begin{equation}
    \vec{u}^\dagger Q_\alpha\vec{u} = \frac{-1}{2} \int \mathrm{d}A |\psi_u(\mathbf{r})|^2 \frac{\mathrm{d}U}{\mathrm{d}\alpha},
    \label{eq:potential}
\end{equation}
where $\vec{u}$ is the vector of modal amplitudes for some general incoming wavefront, $\psi_u(\mathbf{r})$ is the resulting solution to the Helmholtz equation inside the scattering region, and the integral is taken over the whole scattering region. To interpret Eq.~\eqref{eq:potential} in the case of $\alpha = x$, we use $\partial_x S = i(S K_x^\mathrm{in} - K_x^\mathrm{out} S)$ to arrive at a different form of the defining equation of $Q_x$:
\begin{equation}
    Q_x\vec{u} = (K_x^{\text{in}} - S^{\dagger} K_x^{\text{out}} S) \vec{u} = q_x\vec{u},
    \label{eq:momentum}
\end{equation}
where $q_x$ is the eigenvalue. Thus, we can interpret $Q_x$ as describing the momentum difference between the incoming and outgoing wave. We note that this interpretation of Eq.~\ref{eq:momentum} is only true if it describes a single scatterer in an otherwise empty scattering region \cite{ambichl_focusing_2017}. By choosing the eigenvector $\vec{u}_\text{max}$ whose associated eigenvalue $q_x^\text{max}$ is largest, we can perform minimization on it to ensure that the force is minimal for an arbitrary incoming wavefront.

We compute the gradient of $q_x$ by observing how an eigenvalue variation corresponds to a variation in the operator:
\begin{equation}
    \mathrm{d}q_x=\vec{u}^\dagger\mathrm{d}Q_x\vec{u} \label{eq:deval}.
\end{equation}
To vary $Q_x$, we consider Eq.~\eqref{eq:momentum} and arrive at
\begin{equation}
    \mathrm{d}Q_x=-S^{\dagger}{K}_x\mathrm{d}{S}-\mathrm{d}{S}^{\dagger}{K}_x{S},
    \label{eq:dq}
\end{equation}
where we have used ${K}_x^{\text{in}} = {K}_x^{\text{out}} \equiv {K}_x$. To compute $\mathrm{d}S$, we employ a newly introduced technique from Ref.~\cite{horodynski_optimal_2020}. First, we define another GWS operator ${Q}_n$, where our $\alpha$-parameter is now $n$, the index of refraction. We then have
\begin{equation}
    {Q}_n=-i{S}^{\dagger}\frac{\mathrm{d}{S}}{\mathrm{d}n}.
    \label{eq:qn}
\end{equation}
Substituting this new operator into Eq.~\eqref{eq:dq}, we get:
\begin{equation}
    \frac{\mathrm{d}{Q}_x}{\mathrm{d}n} = i({Q}_n{S}^{\dagger} {K}_x {S} - {S}^{\dagger}{K}_x{S}{Q}_n),
    \label{eq:dqdn}
\end{equation}
which we can then take the expectation value of in the $\vec{u}_\text{max}$ eigenstate to get the gradient of this eigenvalue, following Eq.~\eqref{eq:deval}. Why does introducing another operator, $Q_n$, for each design pixel for which we want to change the refractive index, help us in computing the gradient of $q_x$ w.r.t. the refractive index of all design pixels? Ref.~\cite{horodynski_optimal_2020} shows that the elements of $Q_n$ can be computed as 
\begin{equation}
    \left[ Q_n\right]_{ij} = k^2n \int_\mathrm{pixel} \psi_i^*(\mathbf{r}) \psi_j(\mathbf{r}) \mathrm{d}A,
    \label{eq:qnij}
\end{equation}
where the integral is taking over the area of the design pixel, of which we want to compute $\partial_nq_x$, and $\psi_i(\mathbf{r})$ is the wavefunction resulting from an injection of the $i$th mode. Once the wavefunction $\psi_i(\mathbf{r})$ for every mode is computed, we can calculate the integral of Eq.~\eqref{eq:qnij} for every design pixel without requiring the costly solving of the Helmholtz equation. 

It should be noted, however, that this gradient is with respect to the \textit{integrated} refractive index of each design pixel, but we want to be able to change the refractive index continuously across our optimization region. We eventually want to be able to see how $q_x^\text{max}$ reacts to a change in refractive index in a single pixel of our design region, so we will instead limit our integral to a very small square of this region with area $\Delta x \Delta y$.  Over this region, the integral is approximately $|\psi_u(\textbf{r})|^2 k^2 n(\textbf{r}) \Delta x \Delta y$, so we can see that the correct weight factor needed to see how $q_x^\text{max}$ changes when $n$ is changed over a small area is $k^2n(\textbf{r})\Delta x \Delta y$.  Putting this all together, we get the final local gradient of our cost function as:
\begin{equation}
    \frac{\mathrm{d}q_x^\text{max}}{\mathrm{d}n}(\textbf{r})=\vec{u}_\text{max}^\dagger\frac{\mathrm{d}{Q}_x}{\mathrm{d}n}\vec{u}_\text{max}k^2n(\textbf{r})\Delta x \Delta y.
\end{equation}
We now have a cost function (the largest eigenvalue of ${Q}_x$), and its gradient to optimize with.  Our optimization is actually two-fold with every iteration in gradient descent: first, we need to compute the eigenspectrum of ${Q}_x$, and choose the largest eigenvalue, $q_x$.  By doing so, we optimize the incoming wavefront by selecting the mode with the greatest pulling force. Then, we perform the main gradient descent optimization by minimizing this eigenvalue with respect to the refractive index distribution of our design region.  This makes it so that the \textit{strongest possible} pulling force is minimized, so a more general superposition of eigenmodes will have an even weaker pulling force.  This concludes our discussion on the pulling force operator ${Q}_x$, but we are not limited to just looking at the momentum transfer when working in the GWS operator framework.

\subsection{Optical Trapping Stiffness Micromanipulation}

The most widely used method for manipulating objects with light is optical trapping. Trapping stiffness in the $x$-direction is defined as $-\partial_x F_x$, so we define an operator ${W}_x$ that computes the trapping stiffness for an arbitrary incoming wavefront as
\begin{equation}
    {W}_x \equiv -\partial_x Q_x = i(S^\dagger K_x S K_x - K_x S^\dagger K_x S) = i[S^\dagger K_x S, K_x],
    \label{eq:Wx}
\end{equation}
where $[\cdot,\cdot]$ denotes the commutator of two matrices. Again, we define the cost as the largest eigenvalue $w_x^\text{max}$ and use equation \eqref{eq:Wx} to compute the derivative of $W_x$:
\begin{equation}
    \frac{\mathrm{d}{W}_x}{\mathrm{d}n} = \left[\frac{\mathrm{d}}{\mathrm{d}n}\left( S^\dagger K_x S \right),K_x \right] = [[Q_n,S^\dagger K_x S],K_x],
\end{equation}
where we have used the fact that $K_x$ is independent of $n$. Now, the gradient computation follows a similar pattern as for $Q_x$ and is given by 
\begin{equation}
    \frac{\mathrm{d}w_x^\text{max}}{\mathrm{d}n}(\textbf{r})=\vec{v}_\text{max}^\dagger
    \frac{\mathrm{d}{W}_x}{\mathrm{d}n}\vec{v}_\text{max}k^2n({\textbf{r}})\Delta x \Delta y,
\end{equation}
where $\vec{v}_\text{max}$ is the eigenvector of $W_x$ with the largest trapping stiffness.

We employ the same strategy as we did for the $q_x$ optimizations. First, we create a randomize structure, compute the eigenvalues of ${W}_x$, and then minimize the largest eigenvalue with respect to the refractive index distribution while varying the same parameters as before, with one major difference being that we restricted our design region to a circle instead of a square both to highlight the flexibility of our manipulation process and to connect to optical trapping experiments. It should be noted, however, that the maximal eigenstates $\vec{v}_\text{max}$ of $W_x$ do not necessarily result in a stable trapping state. To ensure that the target is stably trapped, we find the appropriate wavestate $\vec{v}$ by considering the following optimization problem
\begin{align}
    \max_{\vec{v} \in \mathbb{C}^n}\quad & \vec{v}^\dagger W_x \vec{v} \\
    \text{s.t.}\quad & \vec{v}^\dagger Q_x \vec{v}=0, \\
    & \vec{v}^\dagger \vec{v} = 1,
\end{align}
This ensures that the force acting on the target is zero at the equilibrium position and that the wavefront is normalized.

\section{Computational Details}

\subsection{Hyperbolic Tangent Regularization}

To create more fabricable structures that only consist of two materials, we use the technique of hyperbolic tangent regularization, where the design region is parametrized by the new regularization variable $t \in (-\infty, \infty)$, instead of directly by the refractive index:
\begin{equation}
    t(n) = \frac{1}{\beta} \text{arctanh} \left(\frac{2(n-n_{\text{min}})} {n_{\text{max}} - n_{\text{min}} }-1 \right),
\end{equation}
where $\beta > 1$ is a parameter that determines how steep the binarization is.  We optimize with this $t$-parameter, and then use the inverse relation to get back to our refractive index distribution after the inverse-designed structure is complete:
\begin{equation}
    n(t)=\frac{(n_{\text{max}}-n_{\text{min}})}{2}(\tanh(\beta t)+1)+n_{\text{min}}.
    \label{eq:tton}
\end{equation}
Following the chain rule, the gradient must be multiplied by a new factor:
\begin{equation}
    \frac{\mathrm{d}n}{\mathrm{d}t}=\frac{\beta(n_{\text{max}}-n_{\text{min}})}{2\cosh^2{(\beta t)}}.
\end{equation}
The procedure is then to start with a randomized $ t$-distribution (rather than a randomized refractive-index distribution), convert it to a refractive-index distribution using Eq.~\eqref{eq:tton}, and perform standard gradient-based optimization. Then, $\beta$ is increased by a factor of 1.07 after every 10 iterations of gradient descent, which forces the $t$ distribution towards $\pm 1$.  When we perform the gradient descent optimization while increasing $\beta$ in this way, the cost function is incentivized to pick structures with a refractive index close to $n_\text{min}$ or $n_\text{max}$. This can be seen from the derivative $\mathrm{d}n/\mathrm{d}t$, in which the rate of change is large near the midpoint of $(n_\text{max}-n_\text{min})/2$, but practically 0 near $n_\text{max}$ and $n_\text{min}$.

\subsection{Parameter Scans}

We will now provide exact details about the nature of our gradient descent optimization for $q_x$ and $w_x$ when varying parameters of the system. We choose to normalize $q_x$ and $w_x$ by the eigenvalues computed for a conventional target to see how well the optimized structure performs when compared to it and to make our cost functions dimensionless. In this paper, we chose a solid structure with a constant refractive index distribution set to $n_\text{max}$ everywhere as our baseline.

When varying a given parameter, we chose default values for the rest of the parameters that gave a good compromise between minimization performance and computation speed for both inverse-design procedures, with exact values found in Table \ref{tab:table}. Each parameter (except $n_\text{max}$, which is dimensionless) is given in arbitrary units. In these units, the radius of the entire simulated region is $R=0.1$, so a target in the center of the scattering region is at a distance of $0.1$ from the wave's source. For all figures in the main paper, we gave $\lambda$ in arbitrary units and the rest of the dimensionful parameters in units of the default value of $\lambda$.  

\begin{table}[t]
    \centering
    \caption{Parameters used in Fig.~2 of the main manuscript.}
    \label{tab:table}
    \begin{tabular}{lccr}
        \toprule % Top horizontal rule
        Parameter & Defaults (tractor beam, trapping)  & Range (tractor beam) & Range (trapping) \\
        Wavelength $\lambda$ (a.u.)& (0.05, 0.0425) & [0.025, 0.075] & [0.02, 0.075] \\
        Refractive index $n_\text{max}$ & (3.0, 3.0) & [2.0, 3.4] & [2.0, 3.4]\\
        Width $\ell$ (a.u.)& (0.02, 0.0325) & [0.01, 0.04] & [0.01, 0.0384]\\
        Distance $d$ (a.u.)& (0.05, 0.1) & [0.05, 0.15] & [0.05, 0.15] \\
    \end{tabular}
\end{table}

To begin our numerical optimization, we partitioned the region into discrete pixels (in this paper, the density was 22,500 pixels per wavelength squared). Next, a structure with a randomized refractive index was generated, and the cost function was defined as the ratio of this new structure's most positive eigenvalue to the base result's. Using this randomized refractive index distribution as an initial guess, gradient descent was performed on the cost function using the SLSQP algorithm. 

Logically, setting up the optimization in this way often results in trivial structures, where all pixels are set to the refractive index of vacuum, i.e., resulting in no structure at all. To fix this, we implemented a constraint to our optimization based on the average ``fill-density'' $\rho$:
\begin{equation}
    \rho \equiv \left< \frac{n(\textbf{r})-n_\text{min}}{n_\text{max}-n_\text{min}} \right>,
\end{equation}
where $\braket{\cdot}$ denotes averaging over the refractive index at every position in the design region. In this paper, we constrained our optimization such that $\rho > 0.5$, i.e., any structure generated must be at least ``half-full'' on average.

\section{Extended data figures}

\begin{figure}[t!]
    \centering
    \includegraphics{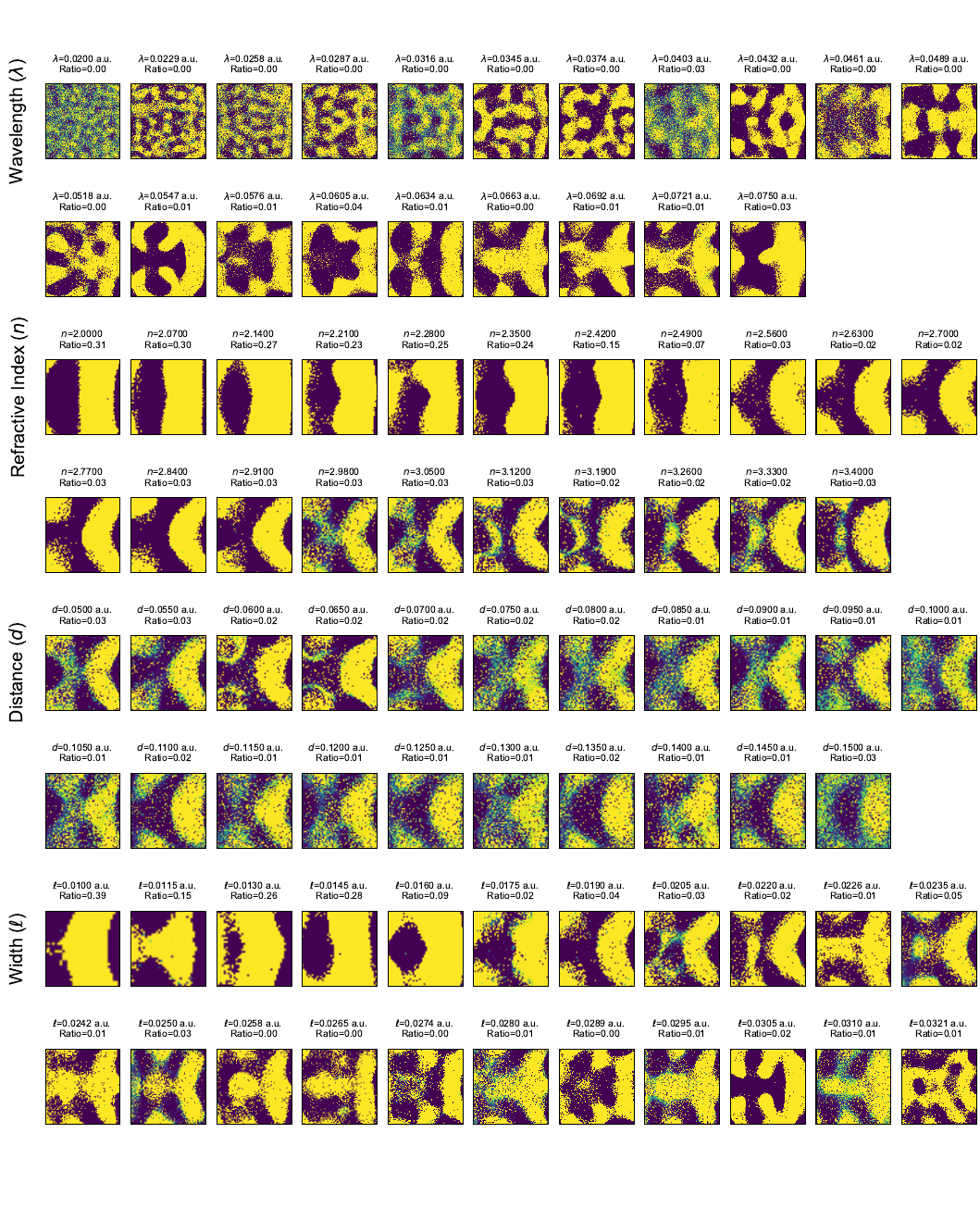}
    \caption{\textbf{Tractor beam minimizing structures.} Plots show all structures used in Fig.~2a-d of the main text with parameters given in arbitrary units. These structures were all generated using tanh regularization to minimize $q_x^\text{max}$. The reduction factor (``ratio'') of the optimal pulling force compared to the conventional structures is also given.}
    \label{fig:3}
\end{figure}

\begin{figure}[t!]
    \centering
    \includegraphics{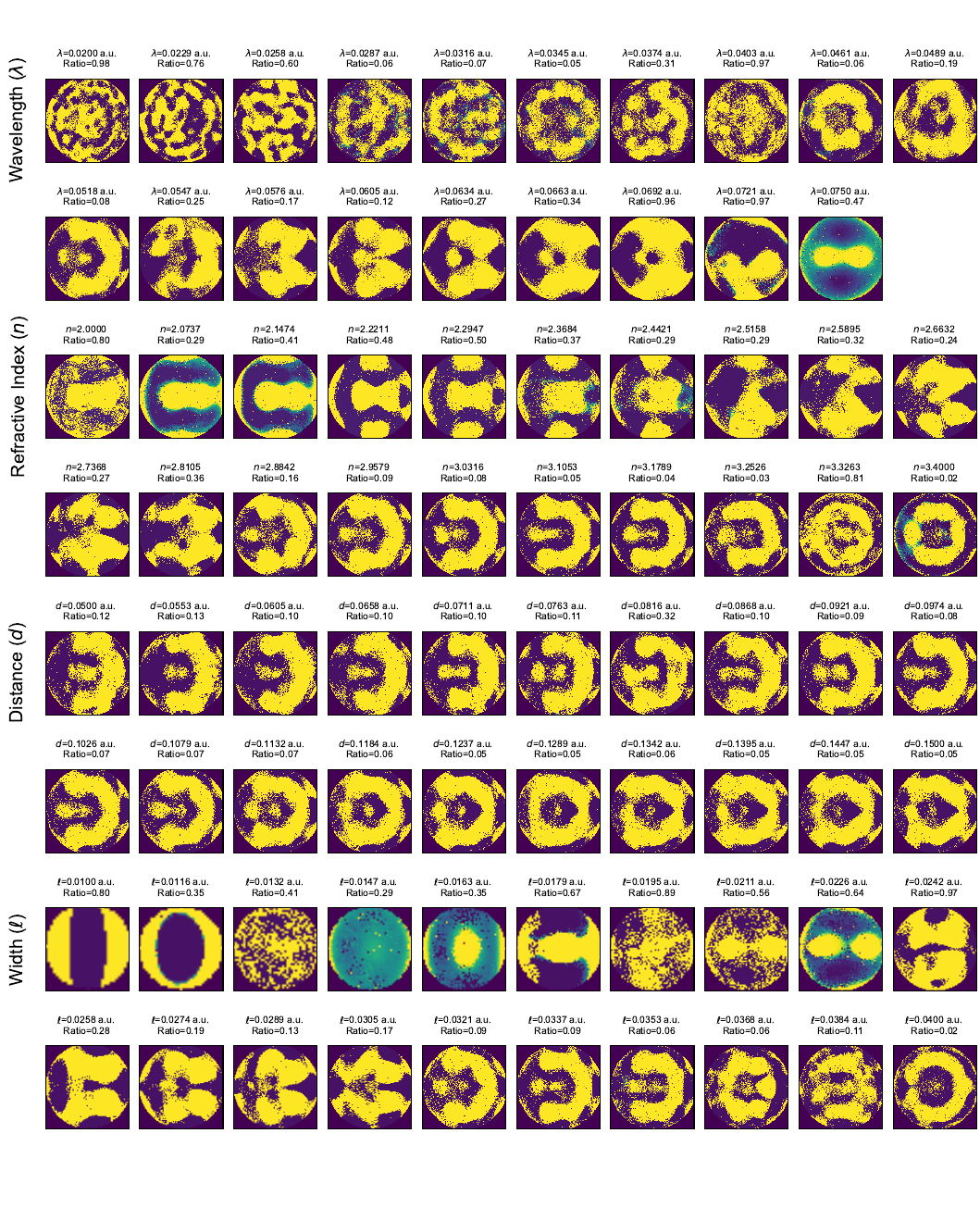}
    \caption{\textbf{Trapping stiffness minimizing structures.} Plots show all structures used in Fig.~2e-h of the main text. With parameters given in arbitrary units. These structures were all generated using tanh regularization to minimize $w_x^\text{max}$. The reduction factor (``ratio'') of the optimal trapping stiffness compared to the conventional structures is also given.}
    \label{fig:3}
\end{figure}

\bibliographystyle{apsrev4-2}
\bibliography{references.bib}